\documentclass[pre,superscriptaddress,twocolumn,footinbib]{revtex4-1}
\pdfoutput=1
\usepackage[colorlinks=true,urlcolor=blue]{hyperref}
\usepackage{amsfonts}
\usepackage{graphicx}
\usepackage{placeins}
\usepackage{amsmath}
\usepackage{xcolor}
\usepackage{color}
\usepackage{bm}
\usepackage[english]{babel}
\usepackage{blindtext}

\definecolor{red}{rgb}{0.75,0,0}
\definecolor{blue}{rgb}{0,0,0.75}
\definecolor{green}{rgb}{0,0.5,0}

\begin{document}

\title{Properties of twisted topological defects in 2D nematic liquid crystals}

\author{D. J. G. Pearce}
\affiliation{Department of Biochemistry, University of Geneva, 1211 Geneva, Switzerland}
\affiliation{Department of Theoretical Physics, University of Geneva, 1211 Geneva, Switzerland}
\affiliation{NCCR Chemical Biology, University of Geneva, 1211 Geneva, Switzerland}
\affiliation{Dept. of Mathematics, Massachusetts Institute of Technology, Massachusetts, United State of America}
\author{K. Kruse}
\affiliation{Department of Biochemistry, University of Geneva, 1211 Geneva, Switzerland}
\affiliation{Department of Theoretical Physics, University of Geneva, 1211 Geneva, Switzerland}
\affiliation{NCCR Chemical Biology, University of Geneva, 1211 Geneva, Switzerland}
\begin{abstract}

Topological defects are one of the most conspicuous features of liquid crystals. In two dimensional nematics, they have been shown to behave effectively as particles with both, charge and orientation, which dictate their interactions. Here, we study  ``twisted'' defects that have a radially dependent orientation. We find that twist can be partially relaxed through the creation and annihilation of defect pairs. By solving the equations for defect motion and calculating the forces on defects, we identify four distinct elements that govern the relative relaxational motion of interacting topological defects, namely attraction, repulsion, co-rotation and co-translation. The interaction of these effects can lead to intricate defect trajectories, which can be controlled by setting relevant timescales. 

\end{abstract}

\maketitle

\section{Introduction}

Topological defects are of great interest in many areas of physics, with prominent examples in cosmology, astrophysics, and crystal structures \cite{Chaikin:1995}. Their existence can be hugely influential on the behaviour of a system, notably for phase transitions in condensed matter~\cite{Kosterlitz:1973}. Beyond the physical context, topological defects are now also recognized to be an important aspect of morphogenesis in biological systems \cite{Singer:2016,Saw:2017gn,Kawaguchi:2017,Maroudas:2021,Guillamat:2021}. Topological defects are a prominent feature of liquid crystals, fluid materials in which the components have a broken rotational symmetry \cite{Frank:1958,deGennes:1995,Kleman:2003}. Indeed their identification lead to the discovery of nematic phases, where the broken rotational symmetry is captured by a director field. Furthermore, topological defects can be used to determine the material properties of nematic materials \cite{Hudson:1989dd,Brugues:2008hx,Zhang:2017,Blanch:2021,Blanch:2021a}.

In two-dimensional nematic liquid crystals, topological defects are point-like disclinations \cite{deGennes:1995,Chaikin:1995,Kleman:2003} with a well-defined half-integer charge and orientation \cite{Vromans:2016,Tang:2017}. These defects have an energetic cost as they disturb the order of the liquid crystal resulting in a Coulomb-like interaction potential between defects~\cite{deGennes:1995,Chaikin:1995,Kleman:2003}. This results in opposite signed topological defects being attracted to each other and eventually annihilating. The situation is more interesting in ``active'' liquid crystals, where the continuous insertion of mechanical stress keeps the system out of equilibrium \cite{Ramaswamy:2010,Marchetti:2013}. Active liquid crystals can exhibit a state referred to as ``active turbulence'' featuring the spontaneous generation of flows along with the proliferation of topological defects \cite{Sanchez:2012}. These active turbulent states generally achieve a steady state density of defects when the elastic and active forces balance \cite{Giomi:2013,Giomi:2014,Giomi:2015,Guillamat:2017}. Furthermore, the interaction between defects has been reported to lead to defect ordered states \cite{DeCamp:2015,Putzig:2016,Doostmohammadi:2016,Oza:2016,Pearce:2019,Shankar:2019,Pearce:2020,Thijssena:2020,Pearce:2021}.

Elastic interactions between defects are present in both, active and passive nematics, and have been shown to play an important role in the torques defects exert on each other \cite{Vromans:2016,Tang:2017,Pearce:2021}. Vromans \& Giomi \cite{Vromans:2016} introduced an efficient method for identifying the orientation of half-integer topological defects in nematic liquid crystals, along with describing their interaction energy, results which have since been expanded upon by Tang \& Selinger \cite{Tang:2017}. In particular, these studies identify an elastic energy dependence on the relative orientation of topological defects, which also results in defects taking curved paths when repelling or attracting one another \cite{Vromans:2016,Tang:2017}. 

In this manuscript we investigate the behaviour of twisted topological defects in two-dimensional nematics. We call a topological defect twisted if the corresponding director field depends on the distance from the defect centre. After characterising individual twisted defects, we extend our analysis to the interaction between two twisted defects. Importantly, the corresponding elastic energy is not necessarily periodic as a function of the relative defect orientation. Instead, it can increase as the square of the relative defect orientation as defects are rotated relative to each other. Furthermore, we identify four distinct behaviours in the motion of pairs of twisted defects, namely the combination of attraction and repulsion with co-translation and co-rotation. We show that the origin of these behaviours lies in the combination of the signs of the defect charges and their twist relative to the background nematic texture. These features can lead to complex relaxation dynamics that can be controlled by temperature as we illustrate with an example of four defects. 

\section{Defects out of equilibrium}

A liquid crystal can be described by the director field $\mathbf{n} = (\cos(\theta),\sin(\theta))$. The director denotes the local average orientation of the anisotropic molecules, where $\theta$ is the angle relative to some background frame of reference. In the case of a nematic, the molecules are elongated and identical under reversal, thus any physics must be invariant under the transformation $\mathbf{n} \to -\mathbf{n}$.

The energy of a nematic liquid crystal depends on variations of the director in space, which in two dimensions can take the form of bend or  splay, each with their own energetic cost. For simplicity, we will consider the limit where the corresponding elastic constants are equal. Then, the total energy of a nematic texture is given by the Frank free energy \cite{deGennes:1995,Chaikin:1995,Kleman:2003}
\begin{equation}
\label{eq:Frankenergy}
E_{\rm{F}} = \frac{K}{2}\int|\nabla \theta|^2\rm{d}A,
\end{equation}
where $K$ is the single elastic constant. 

Topological defects in two dimensional nematics are points, such that the director winds by a multiple of $\pi$ on a closed path encircling them, i.e. $\oint\rm{d}\theta = 2\pi k$ \cite{deGennes:1995,Chaikin:1995,Kleman:2003}. Here $k\in\frac{1}{2}\mathbb{Z}$ is referred to as the topological charge of the defect. At the center (or core) of a defect, the director is not well defined. For an isolated defect with charge $k$, the director field minimising the Frank free energy is 
\begin{equation}
\label{eq:ndef}
\theta = k\phi + \theta_0.
\end{equation}
Here $\phi$ is the polar angle around the core and the phase, $\theta_0$, is a constant that depends on the reference frame.  Note that $\theta$ is independent of the distance $r$ from the defect core. In the case of half integer charge, $k=\pm1/2$, the defect is not rotationally symmetric and the phase $\theta_0$ is related to the defect orientation \cite{Vromans:2016,Tang:2017}.

For a configuration containing multiple defects, the director field minimising the Frank free energy is given by
\begin{equation}
\label{eq:mdef}
\theta = \sum_i k_i\phi_i + \Theta_0.
\end{equation}
Where $\phi_i$ is the polar angle around the core of defect $i$ with charge $k_i$ and $\Theta_0$ is the global phase~\cite{Chaikin:1995}. The single global phase controls the phase at the core of every defect simultaneously, thus Eq.~\ref{eq:mdef} only describes the nematic texture around a set of defects which are in phase with each other. While this is usually sufficient in describing passive liquid crystals, in which the nematic quickly adopts the minimum energy configuration where defects are in phase with each other, this is not necessarily the case in out of equilibrium scenarios, such as those commonly found in active nematics. 

\begin{figure}
	\centering
	\includegraphics[width=\columnwidth]{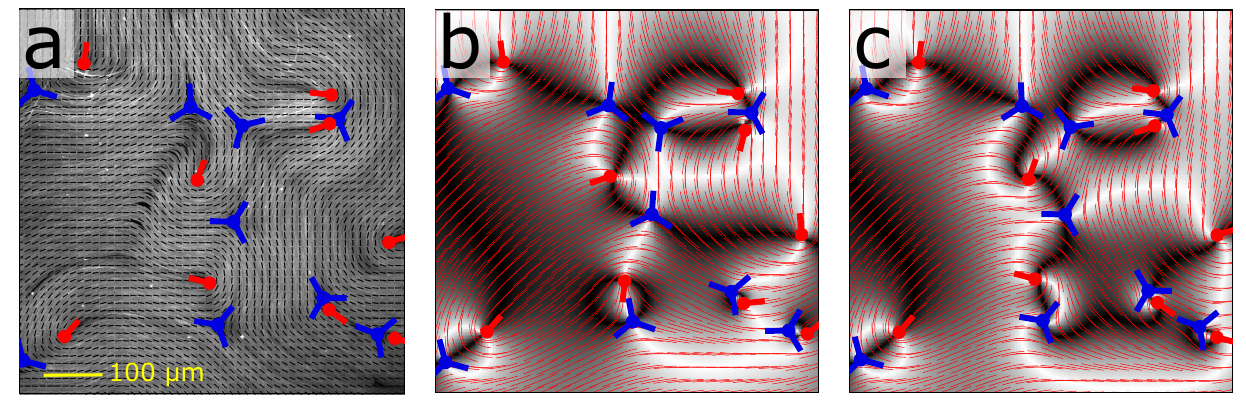}
	\caption{\label{fig:fit} Defects in a nematic out of thermodynamic equilibrium. (a) Snapshot of a microtubule-based active nematic featuring multiple positive (thick red dashes) and negative (blue triads) topological defects. From \cite{Pearce:2021}. (b) Equilibrium nematic texture generated around the positions of the defects in the active nematic by Eq.~\eqref{eq:ndef}. (c) Nematic texture generated by Eq.~\eqref{eq:mtdef}, which recreates the relative phases of each of the defects.}
\end{figure}

In active nematics, there is a constant injection of mechanical stress at the microscopic scale which leads to a proliferation of defects, see Fig.~\ref{fig:fit}a. Using the positions of the defects observed in Fig.~\ref{fig:fit}a we construct a nematic texture using Eq.~\ref{eq:mdef}. Here we see that the single fitting parameter, $\Theta_0$, is not sufficient to match the observed orientation of every defect, see Fig.~\ref{fig:fit}b. This is because the active forces drive defects from their equilibrium positions and orientations, leading to many defects out of phase with each other.

In order to capture the situation depicted in Fig.~\ref{fig:fit}a, we introduce the idea of a twisted topological defect, in which the phase is a function of the radius $r$. Its general form is given by 
\begin{equation}
\label{eq:def}
\theta = k\phi + \tau f(r) + \theta_0,
\end{equation}
where $f(r)$ is a monotonic function. Without loss of generality, we set $\partial_rf(r) < 0$, $\lim_{r\to\infty} f(r) = 0$ and $f(0) = 1$. The twist amplitude $\tau$ controls the defect orientation at its core relative to the nematic texture at $r\to\infty$ rather than to any external frame of reference. It should be noted here that the introduced twist is not the same as the conventional liquid crystalline twist, $\mathbf{n}\cdot(\nabla\times\mathbf{n})$, found in cholesteric phases, which by construction is zero in two dimensions.

We can describe the nematic texture around a set of twisted defects by writing
\begin{equation}
\label{eq:mtdef}
\theta = \sum_i\left[k_i\phi_i + \tau_i f_i(r_i)\right] + \Theta_0,
\end{equation}
where $\partial_rf_i(r) < 0$, $\lim_{r\to\infty} f_i(r) = 0$, and $f_i(0)=1$ for all $i$. With this description, we can control the phase of each defect individually by adjusting the values of $\tau_i$ and $f_i(r)$. To analyze the nematic order field of Fig.~\ref{fig:fit}a, we use a single functional form of $f_i$ for all $i$ to obtain a unique set of values $\left\{\tau_i\right\}$. Explicitly, $f_i(r) = \exp(-10r/L)$, where $L$ is the size of the quadratic domain. Introducing both, the positions and the orientations of the defects in Fig.~\ref{fig:fit}a, into Eq.~\eqref{eq:mtdef} generates the nematic texture given in Fig.~\ref{fig:fit}c, which clearly differs from the one in Fig.~\ref{fig:fit}b and is much closer to the experimental data.

\section{Single twisted defect}

In order to explore the elastic effects of the defect twists introduced in Eq.~\ref{eq:def}, we shall look at simple configurations of defects with high values of twist, starting with a single defect.

\begin{figure}
	\centering
	\includegraphics[width=\columnwidth]{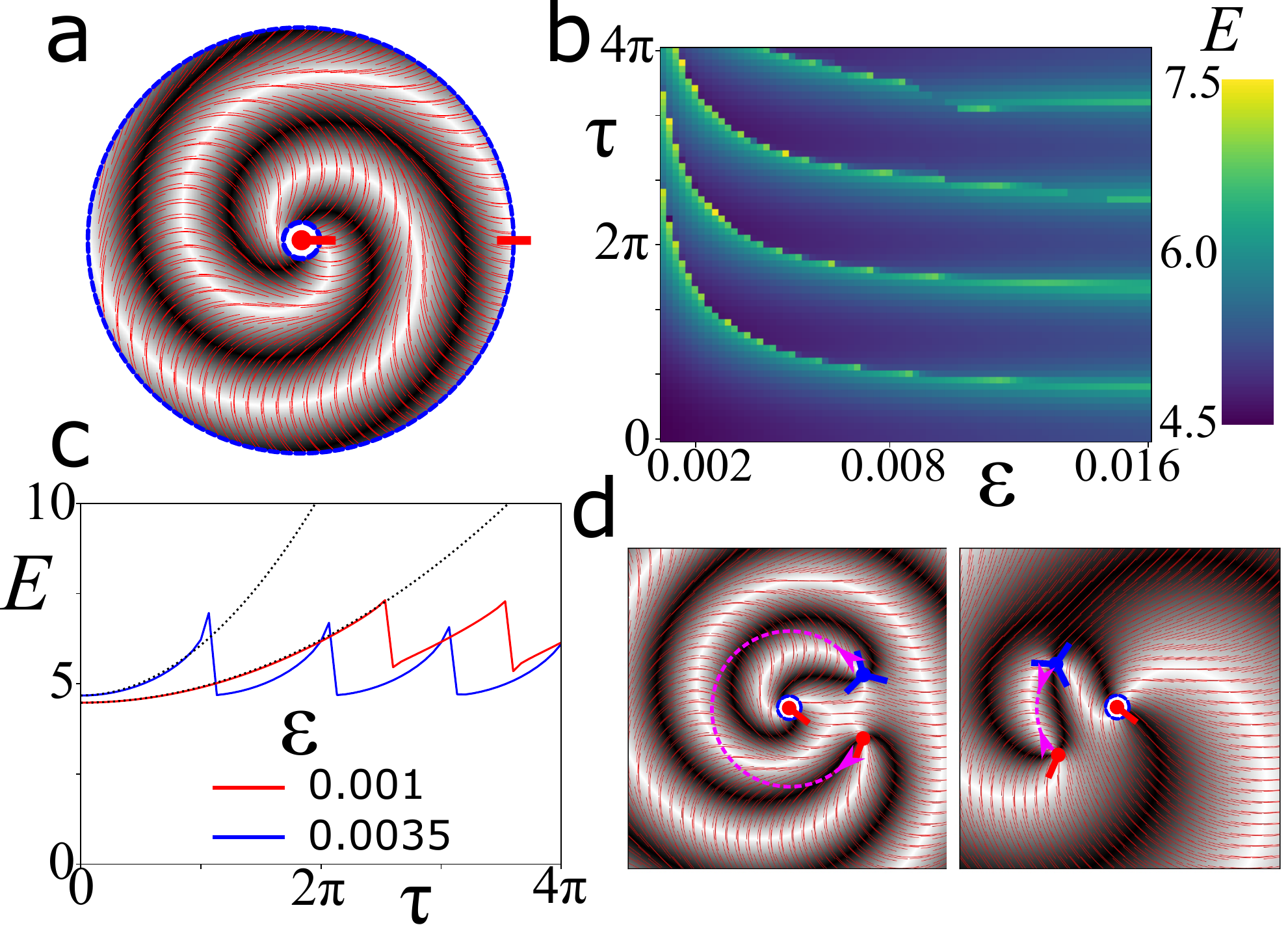}
	\caption{\label{fig:Fig1} Single twisted defects. a) Director field (thin red lines) and corresponding Schlieren texture (grey) of a $+1/2$ defect given by Eqs.~\eqref{eq:def} and \eqref{eq:lin} with $r_0/L = 5/512$, $r_1 = 250/512$, and $\tau = \pi$. Thick red dashes indicate the defect orientation at the fixed boundaries (blue dashed lines). b) Landau-De Gennes energy $E_{\rm{LdG}}$ of nematic textures initialised as in (a) as a function of the twist amplitude $\tau$ and the defect core radius $\epsilon$. c) $E_{\rm{LdG}}$ as a function of $\tau$ for two values of $\epsilon$. Dotted lines indicate $E\sim\tau^2$. d) Schematic of the unzipping removing a $\pi$ twist between the boundaries. The defects nucleate close to the inner fixed boundary (blue dashed lines) and follow pink lines removing $\pi$ twist and annihilating.}
\end{figure}

The nematic texture around a single twisted defect does not minimise the Frank free energy, leading to an increase in the energy by $\Delta E_F = \pi K\tau^2\int f'(r)^2r\rm{dr}$ compared to the defect \eqref{eq:ndef}. Therefore, if the nematic texture around a defect is allowed to relax, in the long time limit $f(r) = \rm{const}$. However, when fixing the phase along two boundaries at different radii, see Fig.~\ref{fig:Fig1}a (blue lines), the twist might not fully relax, resulting in a frustrated configuration, see supp. mov.~1.

When exploring the behaviour of nematics numerically, it is advantageous to work with the nematic tensor $Q_{ij} = S(n_in_j-\delta_{ij}/2)$, where $S$ is the order parameter, instead of the director $\mathbf{n}$ \cite{deGennes:1995,Kleman:2003}. In contrast to the director $\mathbf{n}$, the tensor $\mathsf{Q}$ is well defined even at the defect core, where the order parameter $S=0$. Then, the Frank energy \eqref{eq:Frankenergy} is replaced by the Landau-De Gennes energy 
\begin{equation}
\label{eq:ELdG}
E_{\rm{LdG}} = \frac{K}{2}\int \left[|\nabla \mathsf{Q}|^2 + \frac{1}{\epsilon^2}\rm{tr}\,\mathsf{Q}^2(\rm{tr}\,\mathsf{Q}^2-1)\right]\rm{d}A.
\end{equation}
The Landau-De Gennes energy introduces an additional parameter $\epsilon$, often referred to as the defect core radius. The parameter $\epsilon$ describes the length scale over with the order parameter $S$ varies. 

The relaxation dynamics of the nematic can then be described by
\begin{equation}
\label{eq:H}
\frac{{\partial}Q_{ij}}{\partial t}= \frac{-1}{\gamma}\frac{\delta E_{\rm{LdG}}}{\delta Q_{ij}} = \frac{-K}{\gamma}\left[\Delta Q_{ij} + \frac{(1-S^2)}{\epsilon^2}Q_{ij} \right],
\end{equation}
where we have introduced the rotational viscosity $\gamma$. We numerically integrate Eq.~\eqref{eq:H} using a finite difference scheme on an $N\times N$ grid with $N=512$. When simulating stationary defects, we employ fixed boundary conditions, $\frac{\partial Q_{ij}}{\partial t}= 0$, indicated by blue dashed lines in Figs.~\ref{fig:Fig1}a,d, \ref{fig:Fig2}a,b,e. At the edges of the simulation we impose Neumann boundary conditions, which leave the director free to rotate at the boundary. In this case, there are no topological restrictions on the net defect charge in the simulation domain as is imposed by the Poincar\'{e}-Hopf theorem in the case of fixed boundary conditions \cite{Chaikin:1995,Vromans:2016}. In all cases studied in this work, the simulation domains are sufficiently large such that defects have at least a distance $L/4$ to the boundaries. In the case of repelling defects, which would eventually approach a boundary, simulations are stopped, when the distance of a defect to the nearest boundary is smaller than $L/4$. We scale all lengths by the size $L$ and time by $K/\gamma L^2$. All energies are presented in units of the elastic constant $K$. In keeping with previous works on the topic \cite{Vromans:2016,Tang:2017}, we have opted to focus on the elastic interactions between defects and to neglect hydrodynamic effects introduced by backflow or active stresses.

We start by examining the relaxation of a twisted nematic liquid crystal between two fixed boundaries at $r=r_0$ and $r=r_1$. The nematic texture was initialised by imposing a director field according to Eq.~\eqref{eq:def} with a linear twist
\begin{align}
\label{eq:lin}
f(r)&=\tau(r_1 - r)/(r_1 - r_0),
\end{align}
giving the configuration shown in Fig.~\ref{fig:Fig1}a. This texture is then allowed to relax according to Eq.~\eqref{eq:H} to a stable configuration. Although many of the nematic configurations studied here may appear extremely twisted, it is worth remembering that all results are scaled by the size of the domain. This is because only the net twist between two boundaries is important. Thus the tightness of the twist and therefore the energy density, can be decreased by simply increasing the distance between the boundaries. Integer defects with a high degree of twist have been generated in experimental systems by applying a rotating magnetic field to the nematic liquid crystal 5CB~\cite{Pieranski:2016} or by exposing nematic colloidal suspensions to a chemical gradient~\cite{Navarro:2014}.

In Figure~\ref{fig:Fig1}b, we present the value of $E_{\rm{LdG}}$ after relaxation as a function of the defect core radius $\epsilon$ and the initially applied twist amplitude $\tau$. As $\tau$ is increased, the energy exhibits distinct discontinuities at locations which depend on $\epsilon$. This is highlighted in Fig.~\ref{fig:Fig1}c displaying the Landau-de Gennes energy of the twisted nematic as a function of $\tau$ for two values of $\epsilon$. The energy increases as $\Delta E\sim \tau^2$ before reaching a critical value, at which point it drops.

These drops are associated with the creation and annihilation of a defect pair that is able to ``unzip'' some of the twist between the two boundaries. Since the orientation at the boundaries is fixed, the difference in orientation between a point at $r=r_0$ and a point at $r=r_1$ has a fixed value. In contrast, in the course of the relaxation process, the  twist between these points can change by increments of $\pi$. This is a consequence of the periodic nature of the nematic director. 

For example, consider the nematic texture shown in Fig.~\ref{fig:Fig1}a. Along a straight radial path from $r=r_0$ to $r_1$, the difference in twist is $\Delta\theta_0 = \tau[f(r_1) - f(r_0)] = -\tau$. If the twist is increased such that $\tau\to\tau+\pi$, the orientations at the boundary have not changed, whereas the nematic texture now stores an additional  twist of $\pi$ along the path. This extra $\pi$ twist can only be released by nucleating a pair of oppositely charged defects, which during the relaxation process trace a path around the central boundary to annihilate with each other, see Fig.~\ref{fig:Fig1}d and supp. mov. 2. This effect has been observed previously for integer defects of liquid crystals in a quasi 2D ``dowser'' state~\cite{Pieranski:2016}. There is an energy barrier for nucleating a defect pair. The height of this barrier depends on the defect core radius $\epsilon$. Only if the energy stored in the twisted nematic is above this nucleation energy, can twist be released through this mechanism. 

The critical value of $\tau$ tells us the maximum twist a nematic texture can elastically tolerate for a given scale. For example, the configuration shown in Fig.~\ref{fig:Fig1}a has a twist of $\tau = \pi$ between two boundaries separated by a distance of $r_1-r_0 = 245/512$. From Fig.~\ref{fig:Fig1}b we see this is stable for values of $\epsilon \lesssim 0.003$. In the microtubule based active nematic, the defect core radius is estimated to be $\epsilon \sim 10~\mu$m~\cite{Pearce:2021}, hence a configuration similar to that shown in Fig.~\ref{fig:Fig1}a would only be stable for systems larger than $L\sim 1.6$~mm. For the nematic liquid crystal 5CB, integer defects with high degrees of twist have been observed at length scales $L\sim 10$~mm~\cite{Pieranski:2016}.

We conclude that when the defect core radius is very small compared to the distance between the two boundaries, it is possible for the nematic texture to store a stable twist greater than $\pi$.

\section{Defect interaction energy}

\begin{figure}
	\centering
	\includegraphics[width=\columnwidth]{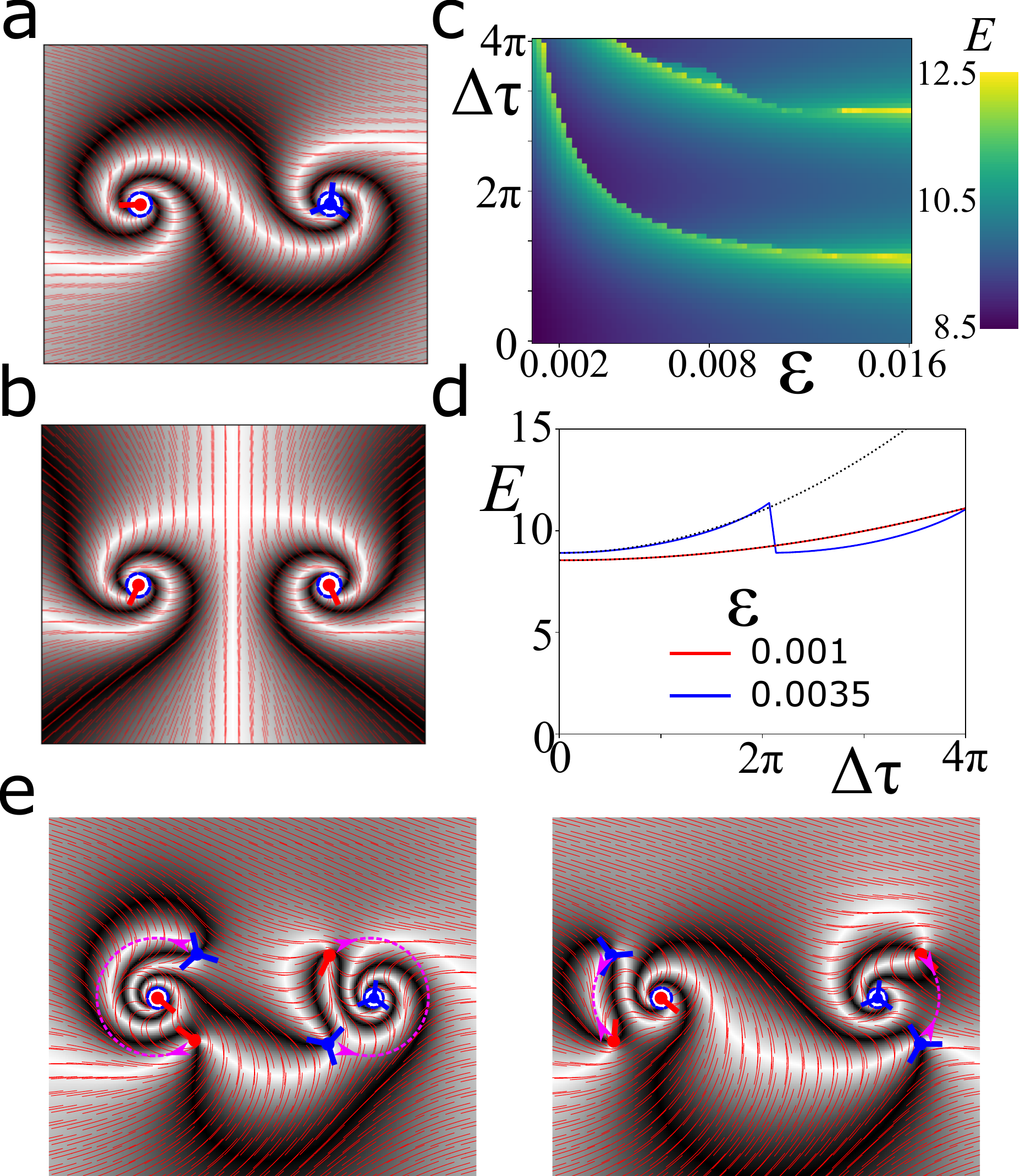}
	\caption{\label{fig:Fig2} Pairs of pinned twisted defects. a) Director field (thin red lines) and corresponding Schlieren texture (grey) of a $+1/2$ and a $-1/2$ defect with twists $\tau_1=-\tau_2=\pi$, see text for details. Thick red dash and blue tripod indicate the defect orientations at the fixed boundaries (blue dashed lines). The defect cores are at $\pm1/8$ along the $x$-axis, $\epsilon_1=\epsilon_2=0.001$, and $r_0/L=5/512$. b) Same as (a), but for two $+1/2$ defects. c) Landau-De Gennes energy $E_{\rm{LdG}}$ of nematic textures initialised as in (a,b) as a function of the twist amplitude $\Delta\tau=\tau_1-\tau_2$ and defect core radius $\epsilon=\epsilon_{1,2}$.  d) $E_{\rm{LdG}}$ as a function of $\Delta\tau$ for two values of $\epsilon$. Dotted lines indicate $E\sim\Delta\tau^2$. e) Schematic of the unzipping process corresponding to a $\pm1/2$ defect pair. A pair of defects is nucleated close to each fixed boundary (blue dashed lines) and follow the pink lines, removing $2\pi$ of twist before annihilating.}
\end{figure}

We now turn our attention to the interaction of defects. Let us consider two fixed circular boundaries with radius $r_0$, each surrounding the core of a defect, and let the angle $\theta$ be given by  
\begin{equation}
\theta = k_1\phi_1 + k_2\phi_2.
\end{equation}
Here, $k_i$, $i=1,2$, are the charges of the two defects and $\phi_i$ are the polar coordinates relative to the respective defects. This creates a nematic texture, which minimises the Frank free energy. Note that the value of $\theta$ is constant along the straight line connecting the two defect cores. We then add a linear twist to each defect according to Eq.~\eqref{eq:lin} with $r_1$ being half the inter-defect distance and $f(r)=0$ for $r>r_1$. This configuration then relaxes according to Eq.~\eqref{eq:H}, while the director is fixed on the boundaries. Examples for the resulting nematic textures are shown in Fig.~\ref{fig:Fig2}a,b and supp. mov.~3. 

In the following, we consider defects with $k=\pm1/2$ and distinguish between defect pairs of opposite, Fig.~\ref{fig:Fig2}a, or the same charge, Fig.~\ref{fig:Fig2}b. In both cases, the defects have equal and opposite twists with respect to the background nematic, $\tau_1 = -\tau_2$. The apparent difference in chirality between the Schlieren textures of these two configurations comes from the fact that $\tau>0$ will rotate the Schlieren texture of opposite charges in opposite directions. 

Two observations should be noted. First, the defects must have an applied twist of opposite sign, $\tau_1 = -\tau_2$. If the defects are twisted with the same sign they remain in phase with each other. In this situation, any twist introduced can be negated by a global rotation of the background nematic, therefore this type of twist is not preserved by the boundaries, see supp. mov.~4. Secondly, in the long time limit only the relative applied twist of the defects is important. Defects can exchange twist by a global rotation of the background nematic and the lowest energy configuration is always that in which the twist is spread evenly between the two boundaries, see supp. mov.~5. Hence, a configuration in which one defect is twisted by $2\pi$ will relax to the configuration in which both defects are twisted by $\pm\pi$.

As in the single defect case, the net difference between the angle $\theta$ for points on the two boundaries is fixed. This implies that the relative twist between the two defects can again exceed $\pi$. After relaxation, the energy is qualitatively similar to the single defect case, Fig.~\ref{fig:Fig2}c. It again features drops in energy associated with an ``unzipping'' of twist by the introduction of pairs of defects, which move around the boundaries and annihilate. Note that the form of the upper line of energy discontinuities results from trapping a defect close to the fixed boundary. Also, the relationship $\Delta E\sim \Delta\tau^2$ is preserved and holds up to the points at which unzipping becomes energetically favourable, Fig.~\ref{fig:Fig2}d. The quadratic dependence on the difference in twist amplitudes is consistent with previous observations made by Vromans \& Giomi \cite{Vromans:2016}. 

Again the critical value of $\tau$ gives us an estimate of the length scale at which a certain phase difference between a pair of defects is stable. The configurations in Fig.~\ref{fig:Fig2}a,b show defects with $\Delta \tau = 2\pi$, which are stable for values of $\epsilon \lesssim 0.003$. Thus to observe a similar configuration in a microtubule based active nematic, the defect separation would need to be of the order of $1$~mm.

During unzipping, defects are nucleated close to the boundaries around the defect cores, where the elastic energy density is highest, Fig.~\ref{fig:Fig2}e. As the twist and the elastic energy associated with this process is equally shared between the two defects, two pairs of defects are nucleated simultaneously during the relaxation process, one close to each boundary. Each new defect pair then encircles the nearby boundary and annihilates, see supp. mov.~6. Compared to the single twisted defect, this process requires double the defect nucleation energy and unzips $2\pi$ of twist. If the separation of the two boundaries is much greater than the defect core radius $\epsilon$, there is not sufficient energy stored in the nematic texture to nucleate defects and a difference in angles between the two boundaries can increase above $2\pi$. Importantly, this implies that the interaction energy between two defects is not necessarily a periodic function of their relative orientation as has been previously predicted \cite{Vromans:2016}. 

\section{Defect Motion}

\begin{figure}
	\centering
	\includegraphics[width=\columnwidth]{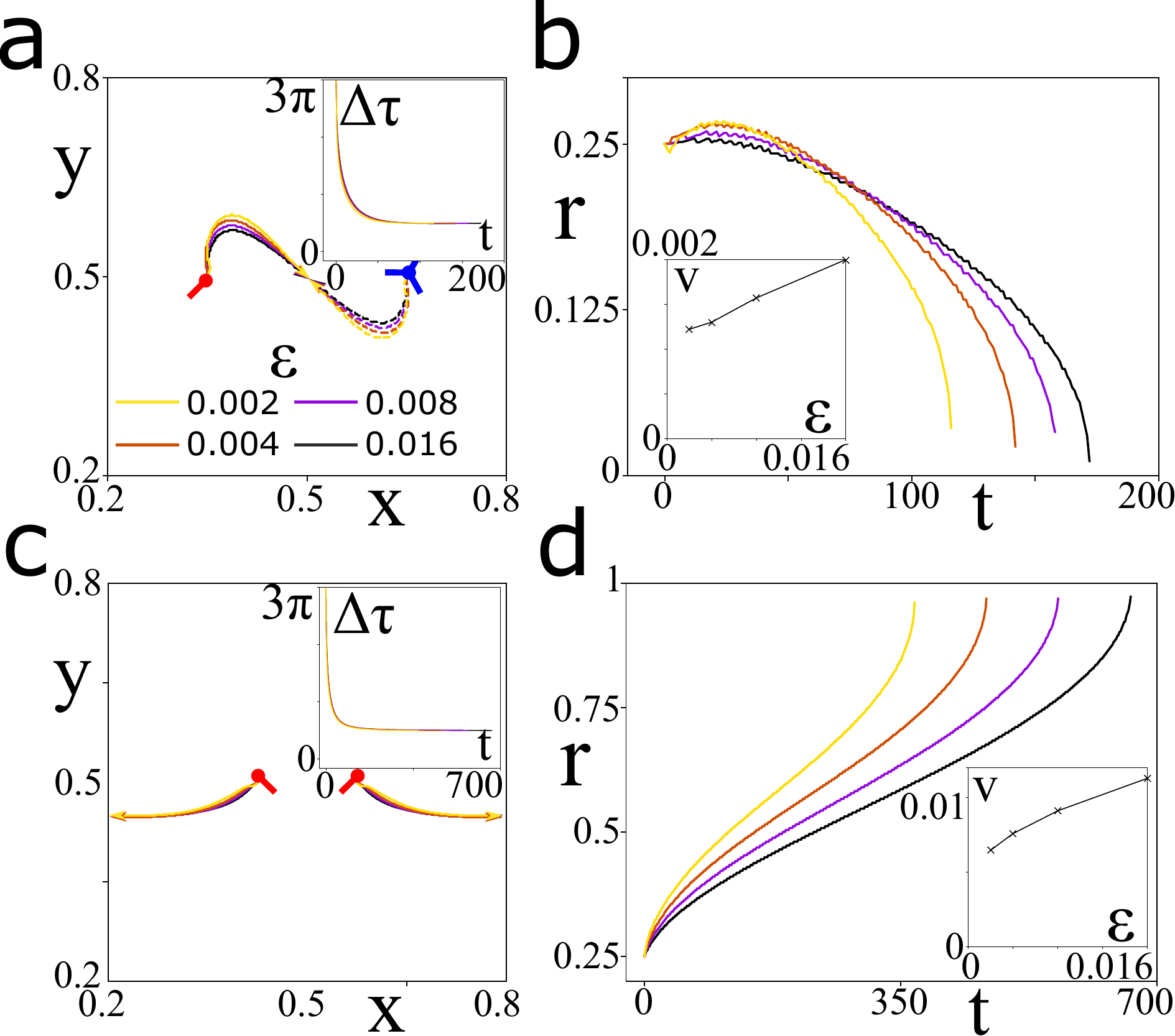}
	\caption{\label{fig:Fig3} Dynamics of out of phase defect pairs for different core radii. a) Defect trajectories for the annihilation of a $\pm1/2$ defect pair initialised with opposite orientations for different values of the core radii $\epsilon$. Inset: The decay of the twist amplitude difference $\Delta\tau$ is independent of $\epsilon$. b) Defect separation as a function of time for the defect pairs shown in (a). Inset: Average velocity of the defects as a function of the defect core radius. c) Same as (a) but for a repelling pair of $+1/2$ defects. d) Same as (b) but for the defect pairs in (c).}
\end{figure}

When the fixed boundaries are removed and the nematic texture is allowed to fully relax to a homogenous state, a $\pm1/2$ defect pair attracts and annihilates, whereas defects with the same sign repel each other. As has been previously reported, when the defects have different orientations, they typically do not travel along a straight line toward or away from each other during the relaxation process \cite{Vromans:2016,Tang:2017}. Instead, defects of opposite signs co-rotate while attracting, Fig.~\ref{fig:Fig3}a,b and supp. mov.~7. This is a direct consequence of the fact that in order to annihilate, two defects must be in phase with each other, i.e. the intermediate texture can be described by Eq.~\ref{eq:mdef}. Thus when two out of phase defects approach each other, they must first remove any relative twist between them before annihilating~\cite{Giomi:2014}. Conversely, defects of the same sign co-translate while repelling each other, Fig.~\ref{fig:Fig3}c,d and supp. mov.~8. Formally, these co-rotations and co-translations are expressed by a rotation of the vector connecting the defect cores and by a translation of their mean position, respectively. 

This behaviour does not change qualitatively with the defect core radius $\epsilon$. Also, the rate at which the twist amplitude difference $\Delta\tau$ is dissipated depends only weakly on the defect core radius $\epsilon$, Fig.~\ref{fig:Fig3}a,c (insets). In contrast, the translational speed of defects exhibits a more pronounced dependence on $\epsilon$, Fig.~\ref{fig:Fig3}b,d, and the velocity of the defects increases with the defect core radius, Fig.~\ref{fig:Fig3}b,d (insets). 

The weak or strong dependence on the defect core radius $\epsilon$ can be explained by the existence of two competing timescales that are relevant for the relaxation dynamics determined by Eq.~\eqref{eq:H}. The timescale associated with relaxation of the nematic director is given by $T_\theta = \Delta r^2\gamma/K$, where $\Delta r$ is the typical length over which $\theta$ varies; in the present case the inter defect spacing. The timescale associated with relaxation of the order parameter is given by $T_S = \epsilon^2\gamma/K$. For the orientation difference between the defects to relax, only the director needs to vary, whereas the order parameter can remain stationary. Hence this process is dictated by $T_\theta$ and therefore independent of $\epsilon$. In contrast, for a defect core to move, both the nematic director and the order parameter must change. This is because the order parameter drops from $S\approx1$ to $S\approx0$ as you approach the defect core. This process thus depends on $T_S$ and hence $\epsilon$.

\begin{figure}
	\centering
	\includegraphics[width=\columnwidth]{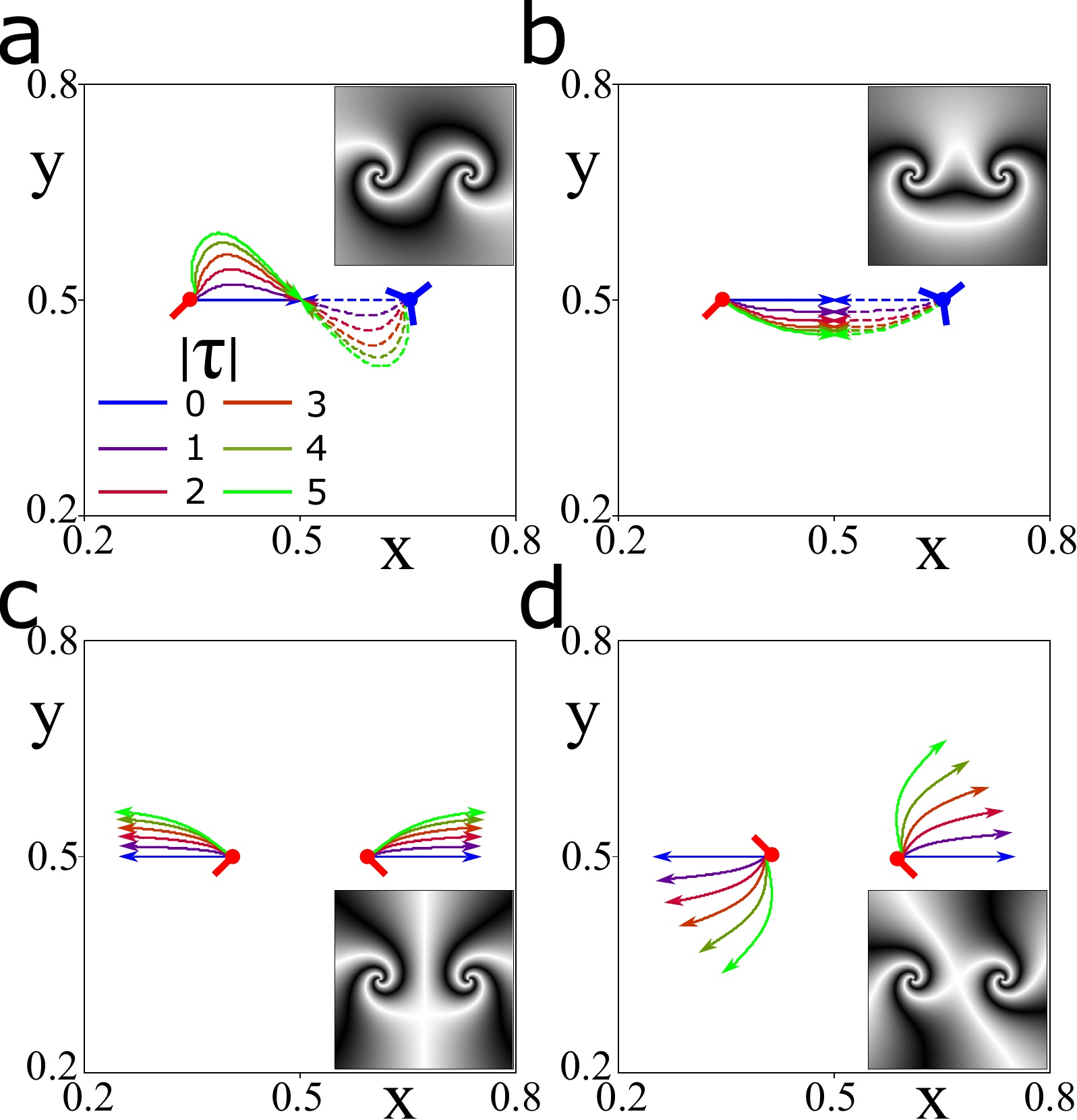}
	\caption{\label{fig:Fig4} Dynamics of out of phase defect pairs for different initial twist amplitude differences following Eq.~\eqref{eq:H}. a) Defects with opposite charges $\pm1/2$ and equal twist amplitude $\tau$; b) defects with opposite charges $\pm1/2$ and opposite twist amplitudes $\pm\tau$; c) defects with equal charges $1/2$ and equal twist amplitudes $\tau$; d) defects with equal charges and opposite twist amplitudes $\pm\tau$. Insets: Schlieren texture of the corresponding initial configurations. Parameter values: $\epsilon=0.008$ and initial $\tau=0$ (blue), 1 (violet), 2 (red), 3 (dark red), 4 (dark green), 5 (bright green). For repelling defects, trajectories are shown up to the point at which they approach the boundary closer than $L/4$ and the twist is almost entirely dissipated. From this point defects move directly away from each other.}
\end{figure}

As mentioned above, for fixed defects, a relative twist can be preserved only if the twists have opposite signs and we only discussed this case. Now that we consider the relaxation dynamics, however, also situations with equal signs of the twist should be analysed. This defines four distinct prototypic scenarios for interacting defect pairs: pairs with like or opposite charges and like or opposite twist amplitudes. The corresponding relaxation dynamics are displayed in Fig.~\ref{fig:Fig4} and supp. mov.~7-10. As above, the dynamics can be either rotational, Fig.~\ref{fig:Fig4}a,d, or translational, Fig.~\ref{fig:Fig4}b,c. This pattern is also apparent from the Schlieren textures of the initial nematic fields, which are either chiral for co-rotating configurations, Fig.~\ref{fig:Fig4}a,d (insets), or achiral for co-translating configurations, Fig.~\ref{fig:Fig4}b,c (insets). By comparing Figs.~\ref{fig:Fig3} and \ref{fig:Fig4} it is clear that the degree of perpendicular motion of the defects depends more strongly on the difference in twist amplitudes $\Delta\tau$ than the size of the defect core radii $\epsilon$.

%only relative phase changes are topologically preserved when equilibrating the intermediate nematic texture between two fixed defects. However, it is possible to initialize the nematic texture between two defects such that they both have a twist relative to the background while remaining in phase with each other. This introduces an additional set of cases; in addition to the defects having equal or opposite topological charge, the defects can have equal or opposite twist relative to the background nematic. This defines four distinct scenarios for interacting defects, the motions of which are displayed in Fig.~\ref{fig:Fig4}. The perpendicular motion of the defects can be sorted into two distinct behaviours. First, co-rotation, in which the mean position of the defects does not change but the vector connecting them rotates, see Fig.~\ref{fig:Fig4}a,d. Secondly, co-translation, in which the vector connecting the defects does not rotate but their mean position translates, see Fig.~\ref{fig:Fig4}b,c. This pattern is also apparent when observing the Schlieren textures of the initial nematic textures which show two distinct patterns either chiral for co-rotating configurations, Fig.~\ref{fig:Fig4}a,d (insets), or achiral for co-translating configurations, Fig.~\ref{fig:Fig4}b,c (insets). By comparing Figs.~\ref{fig:Fig3}\&\ref{fig:Fig4} it is clear that the degree of perpendicular motion of the defects is dependent predominantly on the phase change of the defects rather than the size of the defect core radius.

\section{Forces on the defect cores}

To better understand the motion of defects described in the previous section, we analyze in the following the forces and torques acting on twisted defect cores. In this case, the director field around a set of $N$ twisted defects is conveniently written as
\begin{equation}
\label{eq:multipleDefects}
\theta = \sum_{i}\left[k_i\phi_i + \tau_if_i(r_i)\right],
\end{equation}
where $\phi_i$ and $r_i$ are the polar angle and distance relative to defect $i$, respectively, with $i=1,\ldots,N$. Note that this \textit{ansatz} neglects a possible dependence of $f_i$ on the polar angle $\phi_i$. As above $f_i$ is a monotonic function describing the local twist applied to defect $i$ and $\tau_i$ are the twist amplitudes. 

The interaction between such defects is governed by the Frank energy. It can be obtained by calculating the gradient of $\theta$ around each defect in the polar coordinates centred on that defect with basis set $[\hat{\boldsymbol\phi}_i,\hat{\mathbf{r}}_i]$, such that
\begin{equation}
{\boldsymbol\nabla} \theta = \sum_{i}\left[k_i\hat{\boldsymbol\phi}_i/r_i + \tau_i\partial_{r_i}f_i(r_i)\hat{\mathbf{r}}_i\right].
\end{equation}
We now focus on the case of two interacting defects. The square of the gradient of the angle $\theta$ can then be expressed as
\begin{multline}
|{\boldsymbol\nabla}\theta|^2 = \frac{k_1^2\hat{\boldsymbol\phi}_1^2}{r_1^2} + \frac{k_2^2\hat{\boldsymbol\phi}_2^2}{r_2^2} + \tau_1^2f_1'(r_1)^2\hat{\mathbf{r}}_1^2 + \tau_2^2f_2'(r_2)^2\hat{\mathbf{r}}_2^2\\ + 2\frac{k_1k_2\hat{\boldsymbol\phi}_1\hat{\boldsymbol\phi}_2}{r_1r_2} + 2\tau_1\tau_2f_1'(r_1)f_2'(r_2)\hat{\mathbf{r}}_1\hat{\mathbf{r}}_2\\ + 2\frac{k_1\tau_2f_2'(r_2)\hat{\boldsymbol\phi}_1\hat{\mathbf{r}}_2}{r_1} + 2\frac{k_2\tau_1f_1'(r_1)\hat{\boldsymbol\phi}_2\hat{\mathbf{r}}_1}{r_2}.
\end{multline}
Using the fact that the basis vectors are orthonormal and that $\hat{\boldsymbol\phi}_1\hat{\boldsymbol\phi}_2 = \hat{\mathbf{r}}_1\hat{\mathbf{r}}_2$ and $\hat{\boldsymbol\phi}_1\hat{\mathbf{r}}_2 = -\hat{\mathbf{r}}_1\hat{\boldsymbol\phi}_2$, this can be simplified to
\begin{multline}
\label{eqn:FFE}
|{\boldsymbol\nabla} \theta|^2 = \frac{k_1^2}{r_1^2} + \frac{k_2^2}{r_2^2} + \tau_1^2f_1'(r_1)^2 + \tau_2^2f_2'(r_2)^2\\ + 2\left [\frac{k_1k_2}{r_1r_2} + \tau_1\tau_2f_1'(r_1)f_2'(r_2)\right]\hat{\boldsymbol\phi}_1\hat{\boldsymbol\phi}_2\\ + 2\left [\frac{k_1\tau_2f_2'(r_2)}{r_1} -\frac{k_2\tau_1f_1'(r_1)}{r_2}\right ]\hat{\boldsymbol\phi}_1\hat{\mathbf{r}}_2.
\end{multline}

The first four terms are the defect self-energies associated with the defects' topological charges and twists. The defect self-energy associated with twist is proportional to $\tau^2$, as was observed in Figs.~\ref{fig:Fig1}c and \ref{fig:Fig2}d. This was first observed by Vromans \& Giomi \cite{Vromans:2016} and calculated exactly for a texture minimizing the Frank free energy by Tang \& Selinger \cite{Tang:2017}. The next terms describe the way the two defect charges and twists interact. The term proportional to $\hat{\boldsymbol\phi}_1\hat{\boldsymbol\phi}_2$ has a part proportional to the product of the defect charges that is akin to the Coulomb interaction between electrical charges. It is familiar from interacting defects in the absence of twist, $\tau=0$~\cite{Chaikin:1995,deGennes:1995,Kleman:2003}. The contribution describing the interaction of the two twists is similar, however, the dependence on the radial coordinates is given by $f'$ instead of $1/r$. The final term results from a coupling between twist and charge.

Let us examine the symmetries of the various interaction terms. To this end we assume without loss of generality that both defects are positioned at $\pm r$ along the $x$-axis. We will also assume that they have equal or opposite charges, $k_1=\pm k_2$, as well as equal amplitudes, $|\tau_1| = |\tau_2|$, and distributions, $f_1 = f_2$. The term proportional to $\hat{\boldsymbol\phi}_1\hat{\boldsymbol\phi}_2$ is symmetric for $x\to-x$ and for $y\to-y$. This then describes the energy density between the two defects and can lead to attraction and repulsion.

The final term describes the twist-charge interaction and is proportional to $\hat{\boldsymbol\phi}_1\hat{\boldsymbol r}_2$, which is anti-symmetric in $y$, thus describes the relative energy density above and below the defects. In addition to this, the symmetries of the final term depend on the relative signs of the defect charges and twists. If the defects have either the same charge and same twist or opposite charge and opposite twist, it can be written as $\pm 2k_1\tau_1\left[\frac{f'(r_2)}{r_1}-\frac{f'(r_1)}{r_2}\right]\hat{\boldsymbol\phi}_1\hat{\boldsymbol r}_2$. This expression is anti-symmetric in both $x$ and $y$ and thus can lead to co-rotation of the defects. If in contrast the defects have either the same charge and opposite twist or the opposite charge and the same twist, the final term can be written as $\pm 2k_1\tau_1\left[\frac{f'(r_2)}{r_1}+\frac{f'(r_1)}{r_2}\right]\hat{\boldsymbol\phi}_1\hat{\boldsymbol r}_2$, which is symmetric in $x$ and anti-symmetric in $y$, thus can lead to co-translation. This can be summarized by the rule defects will co-rotate if $\rm{sgn}(k_1k_2) = \rm{sgn}(\tau_1\tau_2)$ and will co-translate if $\rm{sgn}(k_1k_2) = -\rm{sgn}(\tau_1\tau_2)$. Furthermore, the direction of the co-translation or co-rotation depends on the sign of $\tau_1\tau_2$. These different symmetries induce the qualitatively distinct defect dynamics observed in the previous section.

%Since the value of $\hat{\phi}_1\hat{\phi}_2$ is symmetric in $y$, this results in an energy dependent on the relative charge (or twist) and the defect separation. This term leads to the familiar Coulomb like forces observed between defects in phase (i.e. $\tau=0$). The final term is anti-symmetric in $y$, and therefore can lead to vertical forces on the defect cores, thus leading to the off axis motion observed for defect pairs that are twisted relative to the background nematic. If we assume that the magnitude of the twists are equal, $|\tau_1| = |\tau_2|$, and that the twist is equally distributed, $f_1 = f_2$, then there are two distinct scenarios. If the defects have either the same charge and same twist or opposite charge and opposite twist, the final term can be written $\pm 2k_1\tau_1\left[\frac{f'(r_2)}{r_1}-\frac{f'(r_1)}{r_2}\right]$. This is anti-symmetric in $x$ and thus the gradient of this term will have a different sign at the two defect cores. Thus the vertical forces on the defects have different sign. If the defects have either the same charge and opposite twist or the opposite charge and the same twist, the final term can be written $\pm 2k_1\tau_1\left[\frac{f'(r_2)}{r_1}+\frac{f'(r_1)}{r_2}\right]$ which is symmetric in $x$ resulting in a vertical force which has the same sign at both defect cores. This can be summarized by the rule defects will co-rotate if $\rm{sgn}(k_1k_2) = \rm{sgn}(\tau_1\tau_2)$ and will co-translate if $\rm{sgn}(k_1k_2) = -\rm{sgn}(\tau_1\tau_2)$.

To obtain the forces and torques on the defect cores, we differentiate the energy with respect to the defect positions $\underline{x}_i$ and twist amplitudes $\tau_i$. We additionally assume here that the $f_i$ remain fixed such that the time-dependence of all twist is captured by the coefficients $\tau_i$. This allows us to write a set of over-damped dynamical equations for the positions of the defects and their twist amplitudes:
\begin{align}
\label{eq:xdot}
\dot{\underline{x}}_i &= \frac{K\mu}{2} \partial_{\underline{x}_i}\int |\nabla\theta|^2\rm{dA}\\
\label{eq:taudot}
\dot{\tau}_i &= \frac{K\nu}{2} \partial_{\tau_i}\int |\nabla\theta|^2\rm{dA}.
\end{align}
Here, $\mu$ and $\nu$ play the role of a translational and rotational mobility, respectively. We again scale all lengths by the size of the integration domain, $L$, and times by $K\mu/L^2$. The integrals are performed numerically on a $512\times512$ grid.

\begin{figure}
	\centering
	\includegraphics[width=\columnwidth]{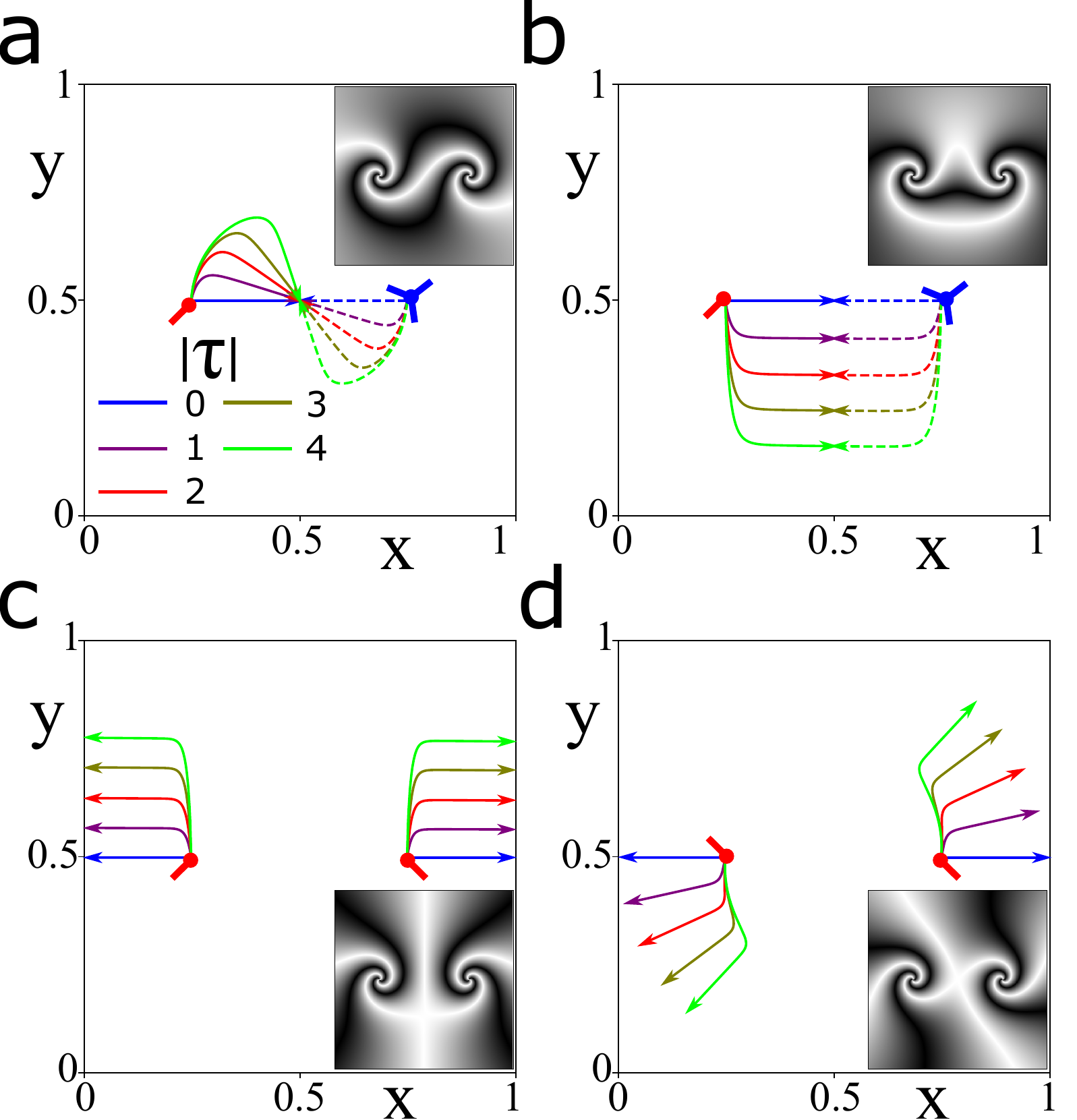}
	\caption{\label{fig:Fig5} Defect trajectories obtained from Eqs.~\eqref{eq:xdot} and \eqref{eq:taudot} corresponding to Fig.~\ref{fig:Fig4}. Parameter values: $\mu/\nu=1.95$, initial $\tau=0$ (blue), 1 (violet), 2 (red), 3 (dark green), 4 (bright green).}
\end{figure}

Let us consider the case $f_i(r_i) = \exp(-r_i/r')$, where $r'$ indicates the length scale over which the twist decays. We numerically integrate these equations to obtain the trajectories of defect pairs for different values of $\tau_i$ and $k_i$, Fig.~\ref{fig:Fig5}. A comparison with Fig.~\ref{fig:Fig4} shows that our model captures the qualitative features of the dynamics of twisted defect pairs. It should be noted here that the dynamics cannot be expected to agree quantitatively as the form $f_i(r_i) = \exp(-r_i/r_0)$ is not a minimiser of the Frank free energy. 

\section{Multi defect dynamics}

\begin{figure}
	\centering
	\includegraphics[width=\columnwidth]{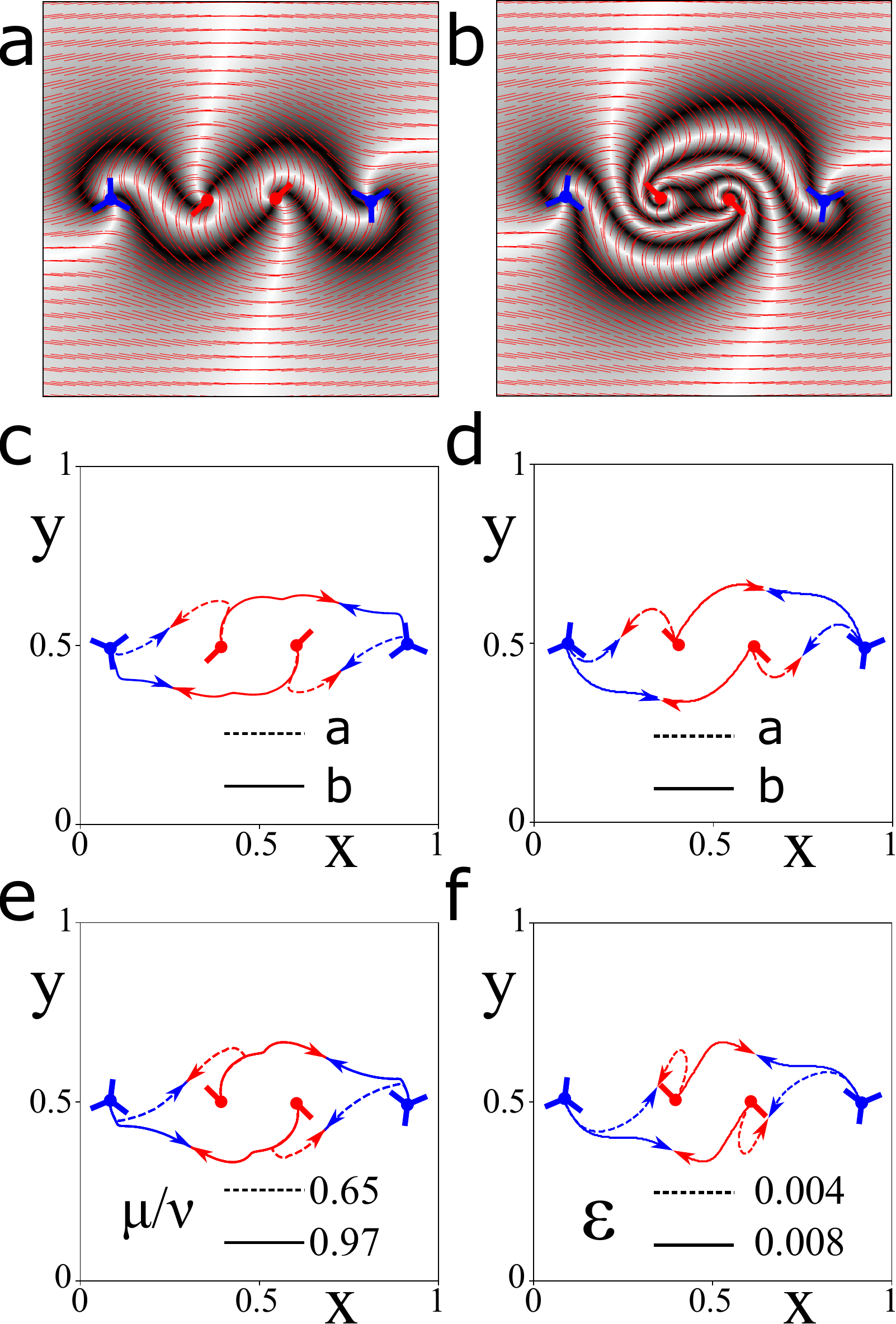}
	\caption{\label{fig:Fig6} Relaxation of two defect pairs. a) Orientation field (red dashes) and corresponding Schlieren texture (grey) given by Eq.~\eqref{eq:multipleDefects} for four defects with charges $-1/2$, $+1/2$, $+1/2$, and $-1/2$ from left to right. The corresponding twist amplitudes are $+\pi/2$, $-\pi/2$, $-\pi/2$ and $+\pi/2$ with $f(r) = \exp(-r_i/r')$ and $r'= 0.1$. b) Same as (a) but with twists $+\pi/2$, $-3\pi/2$, $-3\pi/2$ and $+\pi/2$. c) Defect trajectories corresponding to nematic texture shown in (a) dashed lines and (b) solid lines from Eqs.~\eqref{eq:xdot} and \eqref{eq:taudot}. Parameter values $\mu/\nu = 1.3$. d) Defect trajectories corresponding to (a) dashed lines and (b) solid lines from Eq.~\eqref{eq:H}. Parameter values: $\epsilon=0.016$. e) Defect trajectories corresponding to nematic texture shown in (b) from Eqs.~\eqref{eq:xdot} and \eqref{eq:taudot}. Parameter values: $\mu/\nu=0.65$ (dashed lines) and $0.97$ (solid lines). f) Defect trajectories corresponding to nematic texture shown in (b) from Eq.~\eqref{eq:H}. Parameter values: $\epsilon=0.004$ (dashed lines) and $\epsilon=0.008$ (solid lines).}
\end{figure}

We now turn our attention to decay dynamics of more complex arrangements of twisted defects and their non-trivial trajectories. Consider an arrangement featuring a horizontal line of four equally spaced defects with charges $-1/2$, $+1/2$, $+1/2$, $-1/2$ as viewed from left to right. In a twist free texture, the defects would obviously annihilate as a pair on the left and a pair on the right. If we now apply twists with amplitudes $+\pi/2$, $-\pi/2$, $-\pi/2$, $+\pi/2$ this will result in the nematic texture shown in Fig.~\ref{fig:Fig6}a. In this configuration, any two adjacent defects will co-rotate and adjacent pairs co-rotate in the opposite direction. When this texture relaxes, the defects take curved paths and annihilate as a pair on the left and a pair on the right, dashed lines Fig.~\ref{fig:Fig6}c,d and supp. mov. 11. If the twist of the central pair is now increased, such that the twists are now given by $+\pi/2$, $-3\pi/2$, $-3\pi/2$, $+\pi/2$, we obtain the texture given in Fig.~\ref{fig:Fig6}b. In this configuration, the central pair of defects co-rotate to such a degree that they will switch places before annihilating with the initially furthest negative defect, solid lines Fig.~\ref{fig:Fig6}c,d and supp. mov. 12.

Since the elastic energy of a twisted defect $E\sim\tau^2$, the additional twist of the central pair of defects will initially decay faster than that of the outer pair of defects. Depending on the values of the mobilities $\mu$ and $\nu$, the defects will either reorient quickly or slowly compared to their translational motion. Hence by changing the relative mobilities, we can control the degree of co-rotation that occurs. When the rotational dynamics are fast, the higher twists decay quickly and the defects annihilate as a pair on the left and a pair on the right, dashed lines Fig.~\ref{fig:Fig6}e. However, when the rotational dynamics are slowed down, the additional twist on the central $+1/2$ defects again causes them to switch places and annihilate with the initially furthest $-1/2$ defects, solid lines Fig.~\ref{fig:Fig6}e.

%
%Figure \ref{fig:Fig6}e shows that  
%
%
%is can lead to non-trivial behaviours in a passive nematic liquid crystal and to qualitatively different routes to the equilibrium state.  
%
%Consider the arrangement shown in Fig.~\ref{fig:Fig6}a, featuring a row of four defects with charges $-1/2$, $+1/2$, $+1/2$, $-1/2$ as viewed from left to right. In a twist free texture, the defects would obviously annihilate as a pair on the left and a pair on the right. However, by giving them twists of $+\tau$, $-3\tau$, $-3\tau$, $+\tau$ and depending on the ratio of the mobilities, the dynamic equations \eqref{eq:xdot} and \eqref{eq:taudot} result in annihilation of the two +1/2 defects with the originally furthest -1/2 defects, Fig.~\ref{fig:Fig6}b. Only if the rotational mobility $\nu$ is sufficiently small compared to the translational mobility $\mu$ the right and left pairs of defects annihilate.

The behaviour of the effective defect dynamics is again paralleled by the solutions to the relaxation dynamics \eqref{eq:H}, Fig~\ref{fig:Fig6}f and supp. mov.~12,13. There, the translational mobility is analogous to decreasing the defect core radius $\epsilon$. Depending on its value, either adjacent pairs annihilate (dashed lines) or the central +1/2 defects move away from the nearest -1/2 defects and annihilate with the originally more distant -1/2 defects (solid lines). Excitingly, this means that the trajectories of relaxing defects can be controlled by adjusting the defect core radius, which in an experimental setting is linked to the temperature of the liquid crystal. 

\section{Conclusions}

Topological defects in two dimensional nematics are often described as particles that exist within the liquid crystal and interact through their charge. More recently, this idea has been expanded to include the fact that defects have both charge and orientation \cite{Vromans:2016,Tang:2017}. The defect charges interact through a Coulomb like interaction, with like charges repelling and opposite charges attracting. Relative defect orientations have been shown to increase the interaction energy of topological defects and cause transverse motion as defects attract or repel \cite{Vromans:2016,Tang:2017,Pearce:2021}. 

In this paper, we have considered the situation where defects are not in phase with each other. In that case, which is common in active nematics far from equilibrium, the nematic orientations corresponding to the defects are incompatible with each other and the intermediate texture does not minimise the Frank free energy. In this case, we found that the nematic order can be captured by a superposition of defects with a radially dependent phase, which takes the form of a local twist to the nematic texture. We have also shown that the local orientation of a defect with respect to an external reference frame is typically not sufficient to characterise the relaxation dynamics of defects. Consequently, the relaxation dynamics depend on global features of the orientation field. For a single defect with fixed orientation angle at the boundaries, twist may persist, because the configuration has insufficient energy to generate a defect pair, which is necessary to ``unzip'' a $\pi$ rotation of the director. These findings extend to situations with several defects. 

In cases, where the orientation angle is not constrained at the boundaries and the nematic texture can relax to a defect-free state, the relaxation dynamics depend on global properties of the orientation field as well as on several time scales. The relative twist and charge interact and cause characteristic off-axis motion. The time scales are associated with changes in orientation that keep the order parameter constant and with defect core displacements, which necessarily involve changes in the order parameter. They are in turn set by two length scales, namely, the characteristic distance between defects and the defect core radius. 

The qualitative dependence of defect motion on the core radius could play a significant role in the dynamics of active nematics. Indeed, active nematics are often associated with a constant density of defects and states with long range defect order have been speculated upon both experimentally and theoretically \cite{DeCamp:2015,Putzig:2016,Doostmohammadi:2016,Oza:2016,Pearce:2019,Shankar:2019,Thijssena:2020,Pearce:2020,Pearce:2021}. The process of spontaneous defect ordering is likely to be strongly affected by the two time scales identified above.

\section*{Appendix}
\setcounter{equation}{0}
\def\theequation{A\arabic{equation}}
\subsection{Fitting defect twists to experimental data.}

In the following, we describe how the twist of topological defects can be determined from experimental data. 
The positions and orientations of defects are identified using the methodology outlined by Vromans et al.~\cite{Vromans:2016}. These are then converted into a position, $\underline{r}_i$, charge, $k_i$, and phase, $\psi_i$ for each defect $i$. In the neighbourhood of defect $i$, the nematic texture can then be approximated by
\begin{equation}
\theta = k_i\phi_i + \psi_i.
\end{equation}
If we equate the nematic texture in the neighbourhood of defect $j$ with the general, twisted nematic texture given in Eq.~\ref{eq:mtdef} we can write
\begin{equation}
k_j\phi_j + \psi_j = \sum_i [k_i\phi_i + \tau_if_i(r_i)] + \Theta_0
\end{equation}

We now introduce the matrices, $\mathsf{\phi}$ and $\mathsf{F}$, that will allow us to compute the twists $\tau_i$. The components of $\mathsf{\phi}$ are $\phi_{ij} = \arctan2(\hat{y}.(\underline{r}_j-\underline{r}_i),\hat{x}.(\underline{r}_j-\underline{r}_i))$ and yield the polar angle of defect $j$ around defect $i$, for $i\neq j$. Furthermore, we set $\phi_{ii} = 0$. In turn, the components of $\mathsf{F}$ are $F_{ij} = f_i(r_{ij})$, where $r_{ij}$ is the distance between the cores of defect $i$ and defect $j$, with $F_{ii} = 1$ as per the boundary conditions introduced in the main text.

Finally, we define $\delta_i$ as the deviation of the local phase at defect $i$ from the equilibrium
\begin{equation}
\delta_j = \psi_j - \sum_{i}k_i\phi_{ij} - \Theta_0 = \sum_i\tau_iF_{ij}.
\end{equation}
Since $F$ is symmetric, this equation can easily be inverted provided a suitable choice of $f_i(r)$. Thus the twists required to reproduce a set of defects with given positions, charges and phases are given by 
\begin{equation}
\tau_i = \sum_j\delta_jF^{-1}_{ij}.
\end{equation}
Since we introduced the same number of constraints, $\psi_i$, as parameters, $\tau_i$, we are left with a single parameter, $\Theta_0$, which can be used to fit the remaining background nematic.

\subsection{Supplementary Movie Captions}

\begin{itemize}
	\item Supp\_Mov\_1.mp4: Relaxation of a single $+1/2$ defect between fixed boundaries with a twist given by Eq.~\ref{eq:lin} and $\tau = 2\pi$. This is the same as is shown in Fig.~\ref{fig:Fig1}a.
	
	\item Supp\_Mov\_2.mp4: Animated schematic of unzipping of twist around a $+1/2$ defect.
	
	\item Supp\_Mov\_3.mp4: Relaxation of twist between two boundaries each containing an oppositely charged defect. Each defect is initialised with a linear twist of $\pi$ with opposite sign, resulting in a net twist between the boundaries of $\Delta\tau = 2\pi$. This twist is topologically preserved.

	\item Supp\_Mov\_4.mp4: Relaxation of twist between two boundaries each containing an oppositely charged defect. Each defect is initialised with a linear twist of $\pi$ with same sign, resulting in a net twist between the boundaries of $\Delta\tau = 0$. This twist is not topologically preserved, and is removed by a global phase change.

	\item Supp\_Mov\_5.mp4: Relaxation of twist between two boundaries each containing an oppositely charged defect. The right hand defect is initialised with a linear twist of $\pi$, resulting in a net twist between the boundaries of $\Delta\tau = \pi$. This twist is topologically preserved, but shares equally between the boundaries, leading to each defect having a twist of $\tau = \pm\pi/2$.
	
	\item Supp\_Mov\_6.mp4: Animated schematic of unzipping of twist around a defect pair.

	\item Supp\_Mov\_7.mp4: Relaxation and annihilation of a pair of oppositely charged twisted defects. The defects are twisted out of phase, i.e. $\tau_1 = -\tau_2$. The pair co-rotate before annihilating.

	\item Supp\_Mov\_8.mp4: Relaxation and repulsion of a pair of twisted $+1/2$ defects. The defects are twisted out of phase, i.e. $\tau_1 = -\tau_2$. The pair co-translate while repelling.

	\item Supp\_Mov\_9.mp4: Relaxation and annihilation of a pair of oppositely charged twisted defects. The defects are twisted while remaining phase, i.e. $\tau_1 = \tau_2$. The pair co-translate before annihilating.

	\item Supp\_Mov\_10.mp4: Relaxation and repulsion of a pair of twisted $+1/2$ defects. The defects are twisted while remaining in phase, i.e. $\tau_1 = \tau_2$. The pair co-rotate while repelling.

	\item Supp\_Mov\_11.mp4: Annihilation of two pairs of oppositely charged defects. From left to right, the charges are $-1/2$, $+1/2$, $+1/2$, $-1/2$, and the initial twists are $\tau = +\pi/2$,  $\tau = -\pi/2$,  $\tau = -\pi/2$,  $\tau = +\pi/2$. Each defect annihilates with the closest oppositely charged defect. This is the same data as shown in Fig.~\ref{fig:Fig6}a,d.

	\item Supp\_Mov\_12.mp4: Annihilation of two pairs of oppositely charged defects. From left to right, the charges are $-1/2$, $+1/2$, $+1/2$, $-1/2$, and the initial twists are $\tau = +\pi/2$,  $\tau = -3\pi/2$,  $\tau = -3\pi/2$,  $\tau = +\pi/2$. The defect core radius is small, $\epsilon = 0.004$, each defect annihilates with the closest oppositely charged defect. This is the same data as shown in Fig.~\ref{fig:Fig6}b,f.

	\item Supp\_Mov\_13.mp4: : Annihilation of two pairs of oppositely charged defects. From left to right, the charges are $-1/2$, $+1/2$, $+1/2$, $-1/2$, and the initial twists are $\tau = +\pi/2$,  $\tau = -3\pi/2$,  $\tau = -3\pi/2$,  $\tau = +\pi/2$. The defect core radius is large, $\epsilon = 0.008$, each defect annihilates with the closest oppositely charged defect. This is the same data as shown in Fig.~\ref{fig:Fig6}b,f.

\end{itemize}

\begin{acknowledgments}
This work was funded by the NCCR for Chemical Biology and the SNSF. We thank Jyothishraj Nambisan for providing us the experimental image of the active nematic shown in Fig.~\ref{fig:fit}a and Pau Guillamat for insightful discussions. 
\end{acknowledgments}

\end{document}